\documentclass[aps,prl,twocolumn,groupedaddress]{revtex4-1}

\newcommand{\ket}[1]{\mbox{$ | #1 \rangle $}}

\begin{document}


\title{Investigating the origin of time with trapped ions}


\author{Serge Massar}
\affiliation{Laboratoire d'Information Quantique CP225, Universit\'e libre de Bruxelles, Av. F. D. Roosevelt 50, B-1050 Bruxelles, Belgium}

\author{Philippe Spindel}
\affiliation{
Service de M\'ecanique et Gravitation, Universit\'e de Mons, Facult\'e des Sciences,
20, Place du Parc, B-7000 Mons, Belgium}

\author{Andr\'{e}s   F. Var\'{o}n}
\affiliation{Department Physik, Naturwissenschaftlich-Technische Fakult\"at, Universit\"at Siegen, 57068 Siegen, Germany}

\author{Christof Wunderlich}
\affiliation{Department Physik, Naturwissenschaftlich-Technische Fakult\"at, Universit\"at Siegen, 57068 Siegen, Germany}


\date{\today}

\begin{abstract}

Even though quantum systems in energy eigenstates do not evolve in time, they can exhibit correlations between internal degrees of freedom in such a way that one of the internal degrees of freedom behaves like a clock variable, and thereby defines an internal time, that parametrises the evolution of the other degrees of freedom. This situation is of great interest in quantum cosmology where the invariance under reparametrisation of time implies that the temporal coordinate dissapears and is replaced by the Wheeler-DeWitt constraint. Here we show that this paradox can be investigated experimentally using
the exquisite control now available on moderate size quantum systems. We describe in detail how to implement such an experimental demonstration using the spin and motional degrees of freedom of a single trapped ion.

\end{abstract}


\maketitle


{\em Introduction.} In classical and quantum mechanics time is an external parameter. However this external, absolute, time is not accessible experimentally. Rather time is measured by clocks, that is dynamical systems whose evolution is related in a simple way to the external absolute time.

In classical mechanics substituting clock time for absolute external time is just a change of variables. Not so in quantum mechanics where novel features appear. First, in quantum theory clocks are affected by quantum fluctuations, have inherent uncertainties, and when a clock and a system interact they necessarily disturb each other\cite{Peres}.
Second, an isolated quantum system in an energy eigenstate is in a stationary state. It does not evolve in terms of external time (except for a physically meaningless overall phase).
However, even in this case internal degrees of freedom can be used as clocks and define an internal time\cite{Peres}. Since the state is stationnary,  this internal time is totally uncorrelated to the external time. In fact, one can argue that since external absolute time is unobservable, the time dependent Schr\"odinger equation is just a mathematical convenience, and that all physical quantities (states, observables) should be time independent\cite{PageWootters}.

This issue reappears more forcibly when one considers the quantisation of gravity. Indeed, classical general relativity is invariant under reparametrization of time and therefore has no preferred time variable. If one tries to formally quantize gravity, the invariance of the theory under reparametrisation of time implies that the temporal coordinate dissappears and is replaced by a constraint equation\cite{DeWitt}.  This constraint equation, the Wheeler-DeWitt equation, is ill defined mathematically because of the appearance of second order functional derivatives. However, it can be used to study how time emerges in quantum gravity. In particular in quantum cosmology  one generally studies mini-superspace models, in which only one, or a few, gravitational degrees of freedom are kept. In this context it has been proposed that some internal degrees of freedom can act as clocks and parametrize the evolution of the other degrees of freedom\cite{B,BV,E}. One then recovers an approximate time dependent Schr\"odinger equation. The clock variable should be as "heavy", i.e. as classical, as possible, in order that it be affected as little as possible by quantum fluctuations and  be as little perturbed as possible by the back action of the other degrees of freedom. For these reasons in quantum cosmology the clock variable is generally taken to be the radius of the universe.

Formally, the problems of defining internal time for stationary solutions of the non relativistic Schr\"odinger equation, and for solutions of the Wheeler-DeWitt constraint in the mini-superspace approximation, are practically identical. (The main difference is that the Wheeler-DeWitt constraint is not positive definite, whereas matter Hamiltonians are). For this reason  the issue of internal versus external time in matter systems can be viewed as a proxy for the more fundamental issue of time in quantum cosmology\cite{Peres,PageWootters,Rovelli}.

Motivated by the above considerations, in the present work we study how the emergence of time in stationnary solutions of the non relativistic Schr\"odinger equation can be studied experimentally. The exquisite control that is now  available over moderate size quantum systems enables an increasing number of foundational questions in quantum mechanics to be investigated experimentally (instead of analytically or numerically). For examples see  the recent review articles \cite{Buluta2009,Johanning2009,Schneider2012,Cirac2012,Jordan2012,Mueller2012}. Here, we focus on trapped ion systems, although other systems such as photons, neutral atoms, or solid state qubits could probably also be used as a quantum simulator. 

{\em Model based on spin and vibrational degrees of freedom.}
We consider the simple model in which the clock is realized by a harmonic oscillator and the other degrees of freedom are realised by an angular momentum degree of freedom (a spin S particle), see \cite{Peres, PageWootters, Rovelli}. The Hamiltonian is
\begin{equation}
H=\omega a^\dagger a + \omega \sum_{m=0}^{2S+1} m\vert m \rangle \langle m \vert\ ,
\label{Hamiltonian}
\end{equation}
where $a^\dagger,a$ are the creation and destruction operators for the harmonic oscillator. We have adjusted the potential of the oscillator and  an external magnetic field that couples to the the spin's magnetic moment and thus lifts its degeneracy, such that the frequencies $\omega$ of the oscillator and spin  are equal.

The harmonic oscillator acts as clock. In classical mechanics time is therefore given by the phase of the oscillator, which can be deduced from the position and momentum through $t= \frac{1}{\omega}\arctan q/p$. However this procedure cannot be applied a quantum model because there does not exist a well defined time operator, equivalent in this case a well defined phase operator.
But if the total excitation number of the oscillator is less than $N$
then we can use the Pegg-Barnett phase states \cite{PB1,PB2,PB3}:
\begin{equation}
\vert \Theta_k \rangle = \frac{1}{\sqrt{N}}\sum_{n=0}^{N-1} e^{-i2\pi nk/N} \vert n \rangle\ , \ k=0,\ldots, N-1\ ,
\end{equation}
where $\vert n \rangle=a^{\dagger n} \vert 0 \rangle / \sqrt{n!}$ are the number states.
The phase  states sum to the identity over the space $a^\dagger a < N$:
\begin{equation}
\sum_{k=0}^{N-1} \vert\Theta_k \rangle \langle\Theta_k \vert = \sum_{n=0}^{N-1} \vert n \rangle \langle n \vert \  .
\end{equation}
We can therefore define a phase operator
\begin{equation}
\Theta= \frac{2 \pi }{N}\sum_{k=0}^{N-1} k  \vert\Theta_k \rangle \langle\Theta_k \vert
\end{equation}
whose measurement yields a discretized approximation of the phase of the harmonic oscillator, and therefore of time.

Consider that the spin is initially in the state $\vert \psi\rangle = \sum_{m=0}^{2S+1} a_m \vert m \rangle$ with $a_m$ arbitrary complex amplitudes.
 If we evolve this state according to the time dependent Schr\"odinger equation we find
\begin{equation}
\vert \psi(t_{ext})\rangle = \sum_{m=0}^{2S+1} a_m e^{-i\omega m t_{ext} }\vert m \rangle
\end{equation}
where $t_{ext}$ denotes the external time. Suppose that the clock is in a phase state $\vert \Theta_k \rangle$ corresponding to, say, $k=0$. The overall state at time $t_{ext}=0$  is $\vert \Psi\rangle= \vert \psi\rangle \vert \Theta_0 \rangle$. If we evolve this state according to the time dependent Schr\"odinger equation, the spin and the oscillator evolve independently:
\begin{equation}
\vert \Psi(t_{ext})\rangle= \left(\sum_{m=0}^{2S+1} e^{-i\omega m t_{ext}} a_m \vert m \rangle\right)\left(\frac{1}{\sqrt{N}} \sum_{n=0}^{N-1} e^{-i\omega n t_{ext}} \vert n \rangle\right)\ .
\end{equation}


Now let us consider a particular energy eigenstate for the system formed by the spin and the clock. We project the state $\vert\Psi (t_{ext})\rangle$ onto the subspace of energy $E=\omega M$, for some integer value of $M$. This yields the stationary entangled state
\begin{equation}
\vert \Psi_M\rangle = \sum_{m=0}^{\min\{2S+1,M\}} a_m \vert m \rangle\vert M-m \rangle\ .
\label{PsiM}
\end{equation}
In order to exhibit the evolution in internal time, we carry out the joint measurement of both the phase operator $\Theta$ on the state $\vert \Psi_M\rangle$ (in order to measure the internal time) and of an operator acting only on the spin degrees of freedom.
Suppose the measurement of the phase operator $\Theta$ yields the
result $\frac{2 \pi k}{N}$. The state of the spin conditional on this measurement outcome is
\begin{equation}
\vert \psi (k)\rangle= \frac{e^{i 2 \pi k M/N} }{\sqrt{N}} \sum_{m=0}^{\min\{2S+1,M\}} a_m e^{-i 2 \pi k m/N} \vert m \rangle\ .
\label{equalTimes}
\end{equation}

The interpretation of this result is that each value of internal time (i.e. of $k$) occurs with equal probability, and that the state of the spin conditional on phase $k$ being measured is identical to the spin having evolved for a time $t_{int}=2\pi k / (\omega N)$, where $t_{int}$ is the internal time.

{\em Experimental implementation using trapped ions.}
Before discussing the details of ion trap experiments, we first review the basic requirements for any experiment that wishes to illustrate  the emergence of time in stationary quantum states, using the above states: 1) One needs a system with two degrees of freedom described by the Hamiltonian eq. (\ref{Hamiltonian}); 
2) At (external) time $t_{ext}=0$ one initialises the system in state $\vert \Psi_M\rangle$; 3) At a later (external) time $t_{ext}$, one carries out a measurement of the operator $A_S\otimes  \vert \Theta_k \rangle \langle \Theta_k \vert $. This measurement is not always easy to realise experimentally. One could also carry out alternative measurements provided they allow one to deduce at least approximately the expectation of $A_S\otimes  \vert k \rangle_{PB} \langle k \vert $ (more on these alternative measurements later).
4) One verifies that the results of the measurement are consistent with the predictions of  eq. (\ref{equalTimes}).   Specifically one wants to check that all values of internal time $t_{int}$ are equally probable, that the spin state has evolved in internal time, and that the results are independent of the external time $t_{ext}$ at which the system is measured.

In order to realise the Hamiltonian eq. (\ref{Hamiltonian}), we consider as spin degree of freedom two hyperfine levels of the ion, and as position degree of freedom the motion along one axis of the trap.
We denote the two hyperfine states by $\vert \uparrow \rangle$ and $\vert \downarrow \rangle$.
By adjusting the magnetic field and/or the trap spring constant, one can adjust the energy hyperfine splitting and/or the trap vibration frequency. In this way one can realise the degenerate Hamiltonian eq. (\ref{Hamiltonian}).

A single trapped ion is first initialized into the  vibrational ground state \cite{Wineland1989}. The vibrational state of motion of the atom can be manipulated with high precision to generate Fock, coherent or squeezed states \cite{Wineland1996}. These techniques can be adapted to produce the state
\begin{equation}
a_\uparrow \vert \uparrow \rangle\vert n\rangle +a_\downarrow \vert \downarrow \rangle\vert n+1\rangle\ ,
\label{supstate}
\end{equation}
where $a_\uparrow, a_\downarrow$ are arbitrary complex numbers satisfying $\vert a_\uparrow\vert^2+\vert a_\downarrow\vert^2=1$.

Measurements of the internal states of  trapped ions are usually carried out by registering resonance fluorescence. Depending on the internal state of the ion it will either scatter light or not when subjected to an appropriate light beam  \cite{Dehmelt1986,Sauter1986,Bergquist1986}.
Thus, one can detect with near unity efficiency whether the ion is in state $\vert \uparrow \rangle$ or $\vert \downarrow \rangle$. One can also detect with near unit efficiency whether the ion is in the ground vibrational state $\vert n=0 \rangle$\cite{Wineland1996}: by tuning the laser to the red sideband, the ground state will not fluoresce, as opposed to any other excited vibrational state. These techniques can also be used to initialise the system (by projection) in the state
\begin{equation}
\vert \downarrow \rangle\vert n=0 \rangle\ .
\label{projstate}
\end{equation}

In order to obtain information of the internal state, we precede the projection onto eq. (\ref{projstate}) by a unitary transformation. Arbitrary unitary transformations can be carried out on the spin degree of freedom.
Displacement and squeezing operations can readily be carried out on the vibrational degree of freedom \cite{Wineland1996}. It is thus possible to measure the probability to be in the state
\begin{equation}
\left( \mu \vert \uparrow \rangle + \nu \vert \downarrow \rangle \right) \otimes D(\alpha)S(z)
\vert n=0 \rangle
\label{eq:superposition}
\end{equation}
where $\mu, \nu$ are arbitrary complex parameters satisfying $\vert \mu\vert^2+\vert \nu\vert^2=1$,
$D(\alpha)=\exp\left(\alpha a^\dagger - \alpha^* a\right)$ is the displacement operator, and
$S(z)=\exp\left( (z a^{\dagger 2}-z^* a^2)/2\right)$
the squeezing operator, 
with $\alpha, z $ arbitrary complex numbers, and $a^\dagger, a$ the creation and destruction operators.
The displacement and squeezing operators are ubiquitous in quantum optics, see e.g. \cite{GerryKnight}.

We now show how by varying the parameters $\mu,\nu,\alpha,z$ one can determine how the spin is evolving in terms of the internal time. There are different possibilities that we sketch. The first one being the projection onto coherent states.
To this end, consider the information that can be deduced by measuring the probability to be in the state
$$\vert \psi_{\mu\nu\alpha}\rangle=\left( \mu \vert \uparrow \rangle + \nu \vert \downarrow \rangle \right)\otimes
\vert \alpha \rangle$$
where $\vert \alpha \rangle=D(\alpha)
\vert n=0 \rangle$ is a coherent state (displaced vacuum state).
Measuring the probability to be in state $\vert \alpha \rangle$ gives us information both on the phase and amplitude of the vibrational degree of freedom. On the other hand the ideal measurement (of the phase states) gives us only information about the phase. We thus expect that projecting onto a coherent state should be less precise than projecting onto a phase state. This is confirmed by calculation. The probability to find the state $\vert \psi_{\mu\nu\alpha}\rangle$ is
\begin{eqnarray}
P(\mu,\nu,\alpha)&=&e^{-|\alpha|^2} \left\vert
\mu^* a_\uparrow \frac{\alpha^{*n}}{\sqrt{n!}} +
\nu^* a_\downarrow \frac{\alpha^{*(n+1)}}{\sqrt{(n+1)!}}
\right\vert^2\nonumber\\
&=&e^{-|\alpha|^2} \frac{\vert \alpha\vert^{2n}}{n!}
\left\vert
\mu^* a_\uparrow  +
\nu^* a_\downarrow e^{-i\theta} \frac{\vert \alpha \vert}{\sqrt{n+1}}
\right\vert^2\ ,
\label{eq:P}
\end{eqnarray}
where $\alpha = e^{i\theta}\vert \alpha \vert$.
We thus find that
all values of the phase $\theta$ of the coherent state are equally probable, and that when phase $\theta$ is measured, it is as if the spin had evolved for internal time $t_{int}=\theta / \omega$. 

Note that the measured value of $\vert \alpha\vert$  will fluctuate: its average is $\mu=\sqrt{n}$ and its standard deviation $\sigma=1/2$. The fluctuations disrupt the measurment
(through the factor $\frac{\vert \alpha \vert}{\sqrt{n+1}}$ on the right hand side). Indeed only when $ \vert \alpha\vert = \sqrt{n+1}$ does $\theta$ act exactly like the external time, as predicted by eq. (\ref{equalTimes})
The effect of these fluctuations  will decrease as the excitation number $n$ of the harmonic oscillator increases. This shows that as the clock variable becomes more and more classical (large $n$), the way it is measured becomes less and less relevant, provided some information about the phase of the clock variable is obtained by the measurement.

Note that in the ideal measurement procedure, one would determine the spin state conditional on the phase of the oscillator. However the constraints in ion systems imply that one  in fact measures the oscillator state conditional on the measured spin state. By using Bayes theorem, one can then recover the time evolution of the spin in terms of the oscillator.
In fact the procedure outlined above 
can be rephrased as a measurement of the positive Q function of the oscillator conditional on the spin being found in the state $\mu \vert \uparrow \rangle + \nu \vert \downarrow \rangle $. By extension one could also measure other quasi probability distributions of the oscillator conditional on the spin being found in state $\mu \vert \uparrow \rangle + \nu \vert \downarrow \rangle $. One possibility would be to determine the Wigner function.

Instead of measuring the whole Wigner function, one can just measure
the expected value of the generalized quadrature
$Y_\theta=(ae^{i\theta}-a^\dag e^{-i\theta})/2i$ for some values of $\theta$. Efficient methods to do this for a trapped ion have been proposed, for example, in
\cite{Wallentowitz1995,Lougovski2006,Bastin2006} and applied to detect motional states correlated with internal states of trapped ions, for example, in \cite{Gerritsma2010}. From these measurements one can deduce whether the oscillator is indeed acting as clock for the spin degree of freedom.

As a concrete proposal for such an experiment we outline its implementation using a single Ytterbium ion trapped in an harmonic potential and exposed to a spatially varying magnetic field.
First, we define two internal levels, namely two hyperfine states 
of $^{171}Yb^+$: $\ket{\downarrow}\equiv\ket{S_{1/2}, F=1, m_F=-1}$ and $\ket{\uparrow}\equiv\ket{S_{1/2}, F=1, m_F=0}$. In the presence of a bias magnetic field $B$, these two levels are no longer degenerate, and their energy splitting $\Delta$ can be controlled by adjusting $B$. The states $\ket{\downarrow}$ and $\ket{\uparrow}$ can be individually addressed using an auxiliary internal state $\ket{aux}\equiv\ket{S_{1/2}, F=0}$ \cite{Piltz2014} separated from levels $\ket{\downarrow}$ and $\ket{\uparrow}$ by about 12.6 GHz. The auxiliary level  is important for the preparation of the initial state and the final read-out. Population transfer between $\ket{aux}$ and any of the levels $\ket{\downarrow}$ and $\ket{\uparrow}$ is done using microwave pulses at the corresponding frequency and polarization. The qubit $\{\ket{\downarrow}, \ket{\uparrow}\}$ can be rotated using either a resonant radio-frequency field or two microwave fields via state \ket{aux}. 

A magnetic field gradient allows to couple vibrational and internal levels using RF or microwave radiation \cite{Mintert2001,Johanning2009a}. This coupling is possible due to the effective Lamb-Dicke parameter $\left|\eta_{eff}\right|=\sqrt{\eta^2+\kappa^2}$. The usual Lamb-Dicke parameter (LDP) $\eta$ can be neglected compared to the gradient depending part $\kappa$
due to the long wavelength of RF and microwave radiation. However, the magnetic field gradient $\partial_z B$ along the $z$ direction gives rise to an effective LDP $\left|\eta_{eff}\right|\simeq \left|\kappa\right|
=\frac{\Delta z g_F \mu_B
\left|\partial_z B \right|}{\hbar\omega}$, where $\Delta z=\sqrt{\hbar/2m\omega}$ is the extension of the vibrational ground state wave function. Due to this effective LDP, resonances between internal states are accompanied by blue and red sideband transitions. Driving these sidebands allows for generating a wide range of effective Hamiltonians coupling internal and vibrational states \cite{Wallentowitz1995,deMatos1998,Lougovski2006,Bastin2006,Franca2007}.

A key factor is the energy splitting $\Delta$ between the levels $\ket{\uparrow}$ and $\ket{\downarrow}$ which needs to be chosen as an integer number of the harmonic trap energy level separation $\hbar \omega$.  We choose $\Delta =\hbar \omega$ making the levels $\ket{\downarrow}\ket{n=1}$ and $\ket{\uparrow}\ket{n=0}$ degenerate.

The proposed experiment consist of three steps. The first one is the preparation of the ion in the superposition state $\frac{1}{\sqrt{2}}\uparrow\ket{\uparrow}\ket{n=0}
+\frac{1}{\sqrt{2}}\downarrow\ket{\downarrow}\ket{n=1}$, where for definitness we have chosen $a_\uparrow=a_\downarrow=\frac{1}{\sqrt{2}}$, and restrict to the $n=0$ and $n=1$ space. The second step is to wait an arbitrary external time and the third step comprises detection of the internal state and the motional state of the ion. This last step has the difficulty that the detection of the internal state, can, depending on its outcome, incoherently modify the motional state of the ion. The detection of the motional state is achieved by first mapping the motional state onto the internal state and then detecting the internal state. Each of these steps is outlined in what follows.

Before preparing the degenerate superposition, the ion is initialized in a well defined vibrational state. This can be done by sideband cooling the ion to the ground state such that the ion is found in state $\ket{aux}\ket{n=0}$. Then a circularly polarized microwave $\pi-$pulse on the blue sideband of the $\ket{aux} - \ket{\downarrow}$ resonance prepares state $\ket{\downarrow}\ket{n=1}$.  A  $\pi/2-$pulse on the red sideband of the $\ket{\downarrow}- \ket{\uparrow}$ finally generates the desired superposition state.

To determine the correlations between the internal clock variable and the other system's degree of freedom, it is necessary to measure two degrees of freedom: the spin state and the motional state. In special cases it is possible to determine both with a single measurement by performing a projection onto the coherent states (eq. \ref{eq:P}). In general, sequential measurements can be carried out. The first measurement is to be done on the spin state followed by a determination of the correlated motional state. 

The detection of the internal state in the $\sigma_z$-basis of the qubit $\{\ket{\downarrow}, \ket{\uparrow}\}$ is carried out by measuring state-selectively scattered resonance fluorescence on the $\ket{S_{1/2}, F=1} \leftrightarrow \ket{ P_{1/2}, F=0}$ resonance \cite{Hannemann2002}. 
The qubit measurement is preceded by a microwave $\pi$-pulse transferring the population from  state $\ket{\downarrow}$ (or $\ket{\uparrow}$) to $\ket{aux}$. Then the absence of resonance fluorescence in the subsequent measurement indicates initial population of $\ket{\downarrow}$ (or $\ket{\uparrow}$). The motional state is not altered during this measurement, since no light was scattered. A detection of the qubit state in an arbitrary basis is attained by an appropriate rotation of the qubit $\{\ket{\downarrow}, \ket{\uparrow}\}$  preceding  the detection process described above \cite{Hannemann2002}.  
  The motional state correlated to either \ket{\downarrow}  or \ket{\uparrow} after such a null detection event is obtained by mapping the motional states onto two  internal states of the ion \cite{Wallentowitz1995,Lougovski2006,Bastin2006,Gerritsma2010,deMatos1998,Franca2007}, such as $\ket{S_{1/2}, F=0}$ and $\ket{S_{1/2}, F=1, m_F=+1}$), and then measuring the internal states 
as described above.  This mapping is achieved by using  red and blue sideband transitions that accompany the resonance between the ion's internal states. 

We presented in detail how an experiment could be realised in the case where the clock variable is the vibrational degree of freedom of the ion, and the internal state is a spin degree of freedom. In future work it would be interesting to consider clock variables that are even closer to those encountered in quantum cosmology. The difficulty is that the clock variable in mini-superspace would have to be modeled as a system with negative kinetic energy. At first sight this seems impossible. It could however possibly be realised in atomic lattices, using ideas borrowed from a recent experiment that demonstrated negative temperatures \cite{BRSHRBS}. This would allow the investigation of many additional phenomena, including the backreaction of the clock on the matter degrees of freedom, and possible violations of unitarity in the evolution of the matter degrees of freedom when the clock is not classical enough.

\begin{acknowledgments}
We thank Marianne Rooman and Fran\c{c}ois Englert for many stimulating discussions, over many years, about the origin of time.
This work has been partially supported by the "Communaut\'e fran\c{c}aise de Belgique -- Actions de Recherche concert\'ees", by IISN-Belgium (convention 4.4511.06), and by the European
Community's Seventh Framework Programme (FP7/2007-2013) under Grant Agreement No. 270843 (iQIT),
\end{acknowledgments}


\end{document}